\documentclass[10pt, conference, compsocconf]{IEEEtran}
\usepackage{booktabs}
\usepackage{cite}

\ifCLASSINFOpdf
   \usepackage[pdftex]{graphicx}
\else
   \usepackage[dvips]{graphicx}
\fi
\usepackage[ruled,norelsize]{algorithm2e}
\usepackage{url}

\usepackage{subcaption}
\usepackage{caption}
\usepackage{color}
\usepackage{eufrak}
\usepackage{color, colortbl}

\begin{document}
%
\title{Applying Private Information Retrieval to Lightweight Bitcoin Clients}



%
\author{\IEEEauthorblockN{Kaihua Qin\IEEEauthorrefmark{1},
Henryk Hadass\IEEEauthorrefmark{1},
Arthur Gervais\IEEEauthorrefmark{1} and
Joel Reardon\IEEEauthorrefmark{2}}
\IEEEauthorblockA{\IEEEauthorrefmark{1}Department of Computing, Imperial College London\\
Email: kaihua.qin@imperial.ac.uk, henryk.hadass14@imperial.ac.uk, a.gervais@imperial.ac.uk}
\IEEEauthorblockA{\IEEEauthorrefmark{2}Department of Computer Science, University of Calgary\\
Email: joel.reardon@ucalgary.ca}}

\maketitle

\begin{abstract}
Lightweight Bitcoin clients execute a Simple Payment Verification (SPV) protocol to verify the validity of transactions related to a particular user. Currently, lightweight clients use Bloom filters to significantly reduce the amount of bandwidth required to validate a particular transaction. This is despite the fact that research has shown that Bloom filters are insufficient at preserving the privacy of clients' queries.

In this paper we describe our design of an SPV protocol that leverages Private Information Retrieval (PIR) to create fully private and performant queries. We show that our protocol has a low bandwidth and latency cost; properties that make our protocol a viable alternative for lightweight Bitcoin clients and other cryptocurrencies with a similar SPV model. In contract to Bloom filters, our PIR-based approach offers deterministic privacy to the user.

Among our results, we show that in the worst case, clients who would like to verify 100 transactions occurring in the past week incurs a bandwidth cost of 33.54 MB with an associated latency of approximately 4.8 minutes, when using our protocol. The same query executed using the Bloom-filter-based SPV protocol incurs a bandwidth cost of 12.85 MB; this is a modest overhead considering the privacy guarantees it provides.

\end{abstract}



%
\IEEEpeerreviewmaketitle

\section{Introduction}

Bitcoin~\cite{nakamoto2008bitcoin}, a pseudonymous cryptocurrency created in
2008, enables users to perform irreversible electronic transactions. Once a
transaction occurs, the Bitcoin peer-to-peer network records this fact in a
replicated and public database called the blockchain. The clients that verify
and store every transaction in the network are called full nodes. This
verification process is resource intensive, requiring hardware and storage
nearly exceeding personal computers. At the time of writing, the Bitcoin
blockchain was approximately $210$ GB in size, monotonically growing every day.

Resource constrained devices, such as mobile phones, are unable to store the
entire contents of the Bitcoin blockchain given its prohibitive size\footnote{At the time of
writing, an iPhone 8 with 256 GB of storage  costs \$749, a Samsung
Galaxy Note 9 with 512 GB of storage costs  \$1,249.99 and a SanDisk
Ultra 400 GB microSD card alone costs \$175.13.}. Moreover,
bandwidth usage on mobile phones is usually metered, which is
another factor preventing full nodes from being run on such devices.
These limitations, combined with the scale of the blockchain,
prevents participants in the system from being able to validate
their own transactions on their mobile phones. Bitcoin, however, was
aimed at completely removing such trust assumptions: no central
third-party should be required for it to work properly. This
suggests that users need to run their own full nodes at home in
order to participate in the network.

This drawback was already anticipated when Bitcoin was first
introduced~\cite{nakamoto2008bitcoin}. Nakamoto suggests that a lightweight version of the full node protocol should be used in resource constrained environments, which was called Simple Payment Verification (SPV). SPV simply verifies
transactions which are relevant to a specific user, in contrast to a full node
which verifies every single transaction that has ever occurred. 

A simple implementation of SPV requires the client to download the whole Bitcoin
blockchain, much like a full node, in order to verify a single transaction. This
approach preserves the privacy of the client since no other network participant
can determine which transaction the client is interested in verifying. However,
this approach wastes bandwidth since the majority of the downloaded data is
discarded. We call this solution Naive SPV.

SPV as described in the Bitcoin white-paper was only realised in 2012 when a
Bloom-filter-based SPV protocol was proposed in Bitcoin Improvement Proposal
number 37 (BIP-37)~\cite{bip37}. This proposal allowed the SPV protocol to
significantly reduce the amount of bandwidth used in verifying a particular
transaction, when compared with Naive SPV\@. This resulted in BIP-37 becoming the de-facto standard in the Bitcoin community for lightweight clients, with mobile cryptocurrency wallet applications such as BitcoinJ~\cite{btcj} implementing it. We call this solution BIP-37 SPV.

Using Bloom filters, however, does not maintain the privacy of a user's queries. This has been acknowledged not only by the authors of BIP-37~\cite{mike}, but also by the Bitcoin community~\cite{jonas} and by researchers in this field such as Gervais et al.~\cite{gervais2014privacy}. Bloom filters present a trade-off between bandwidth and privacy. In the existing Bitcoin wallet implementations, almost no privacy is guaranteed because a low false positive rate is chosen for the sake of bandwidth. Bloom filters are moreover vulnerable to intersection attacks even though sacrificing bandwidth and increasing the false positive rate, which means that the adversary can read the user's interest by intersecting the results from different Bloom filters of the same wallet. Gervais et al.~\cite{gervais2014privacy} offer solutions to improve the privacy provisions of Bloom filters. Nevertheless, we would like to provide the user with full and deterministic privacy.

In a public payments network such as Bitcoin, privacy is a fundamental requirement. Users who do not adequately preserve their privacy allow full node peers to discover the Bitcoin addresses they own and which transactions they have been involved in. Upon discovering this information, adversarial peers can engage in a denial of service to users or addresses which they disfavour. In addition, these peers can engage in a trivial process of tracing the funds of a particular user, given that the Bitcoin blockchain is public. This tracing could expose an individual to personal harm if it was discovered that they have access to a large quantity of Bitcoin. Due to these risks, it is highly desirable that Bitcoin's users preserve their privacy. 



\textit{Summary}: In this paper we describe our design of a Private Information Retrieval (PIR) based SPV client to demonstrate that a fully private, low bandwidth and low latency lightweight Bitcoin client is feasible. The contributions of this work are the following:

\begin{itemize}
  \item \textbf{First PIR system for blockchain:} To the best of our knowledge, we are the first to design a PIR scheme to increase the privacy properties of cryptocurrency based systems.
  \item \textbf{Different PIR schemes:} We show that in the single-server setting our protocol is as efficient as BIP-37 SPV, while in the multi-server setting it has a comparable bandwidth cost.
  \item \textbf{Bandwidth comparison:} We provide a detailed analysis of the bandwidth cost of our novel protocol and compare its cost to the Naive and BIP-37 SPV protocols. We show that in both instances our protocol is far more bandwidth efficient than the Naive SPV protocol.
  \item \textbf{Practical Latency:} We also show that the latency of executing queries in our protocol is tolerable, with clients being able to perform fully private SPV in far less than a minute when verifying recent transactions.
\end{itemize}

This work is organised as follows. Section~\ref{sec:background} covers the necessary background. Section~\ref{sec:systemoverview} provides an overview of our system and Section~\ref{sec:systemdetails} outlines the details. 
In Section~\ref{sec:evaluation}, we present our evaluation. Section~\ref{sec:futurework} covers future work which improves our implementation and in Section~\ref{sec:relatedwork}, we discuss related work in this field. Finally, we conclude in Section~\ref{sec:conclusion}.

\section{Background}
\label{sec:background}
\subsection{Bitcoin}
Other than full nodes and lightweight clients, Bitcoin has another type of network participant called miners. The role of miners is to secure the Bitcoin blockchain through the execution of a Proof-of-Work (PoW) mechanism.

Blocks, which contain one or more transactions, have the PoW mechanism executed
on them by miners. Blocks are said to be mined once a solution to the PoW
mechanism has been found. The first mined block is called the Genesis block.
Each block contains a block header, which is a lightweight summary of the
contents of the block and is $80$ bytes in size. Each Bitcoin transaction
consist of one or more inputs and outputs, with the inputs describing where the
Bitcoin is coming from and the outputs where it is going to. Each transaction
has a unique identifier called a transaction ID (TXID) which is obtained by
applying a hash function to the transaction data.

The TXIDs of all of the transactions contained in a block are used to construct
a Merkle \cite{merkle1980protocols} tree, whose root is included in the block
header. This Merkle tree root serves as a unique identifier for all of the
transactions included in a particular block. Since the outputs of a transaction
state where the Bitcoin is sent to, spending Bitcoin means using the outputs of
some transaction (for which you were the beneficiary) as the inputs to a new
transaction. Since Bitcoin prevents double spending of the currency, this
divides outputs into two categories: those that have been spent---i.e., used as
inputs to a subsequent transaction---and those that are unspent. The latter are
called Unspent Transaction  Outputs (UTXOs) and they sum to the total amount of
Bitcoin in existence.

UTXOs represent the concept of ``having
bitcoin''---insofar that one can sign value transactions that include the UTXOs as inputs.
Each transaction output includes a Bitcoin address of which there are a number
of different types, the most common of which are Pay-to-Public-Key-Hash (P2PKH)
and Pay-to-Script-Hash (P2SH) \cite{blocksci}. P2PKH are
addresses owned by a single entity whereas P2SH can be multi-signature
addresses. These addresses are usually encoded in the hexadecimal base58
format. Base58 removes characters from its encoding set which are
similar to each other, such as 0, O, I and l, to reduce the risk of
users mixing up distinct characters with similar glyphs.

\subsection{Simple Payment Verification (SPV)}
\label{sec:spv}
Three items are required to perform SPV on a transaction. The transaction itself, the list of TXIDs from the corresponding block and a list of all the block headers from the Genesis block, up to and including the block in which the transaction of interest is included. The transaction is used to calculate its TXID, to check that this TXID exists in the list of TXIDs from the relevant block and to confirm that an addresses belonging to the client is referenced in an output. The list of TXIDs is used to construct the Merkle tree and subsequently calculate the Merkle tree root and check if it matches the value stored in the corresponding block header. The list of block headers is used to prove that the header belonging to the block of interest can be placed in a valid location in the Bitcoin blockchain by recursively hashing them and confirming their corresponding Proofs-of-Work. Once the validity of a transaction has been determined, using the three items outlined above, confidence in the irreversibility of that transaction needs to be established. This is done by ensuring that the block containing the transaction of interest 
is embedded by normally $1$ to $6$ subsequently mined blocks depending on the transaction value~\cite{gervais2016security}.


A naive implementation of an SPV client would first establish one or more connections to a set of full nodes. Next, in order to verify a transaction, the client would request for every single block of transactions from the Genesis block up to the block in which the transaction is included, from the peers it is connected to. For each received block the client would then select the data detailed above and discard the rest. The advantage of this approach is that the full node peers are not able to determine which particular transaction the client is interested in verifying, thus preserving the user's privacy. This is because the SPV client needs to download essentially the whole blockchain for each transaction which it would like to verify. However, this means that this approach is bandwidth intensive since the majority of the downloaded data is not used and is discarded.

\subsection{BIP-37: Bloom filters \& Merkle blocks}

In order to improve bandwidth efficiency over the Naive SPV protocol, the BIP-37 SPV protocol introduced Bloom filters and a corresponding Merkle block to significantly reduce the bandwidth required to verify a particular transaction. We now explain both of these constructions.

A Bloom filter is a space efficient, probabilistic data structure that allows for the testing of an elements' membership in a set. Bloom filters work probabilistically and, by design, only have false positives (FP) but not false negatives (FN).
This means that
Bloom filters reliably tell when an element is \emph{definitely} not in a set, but only offer probabilistic guidance as to whether an element may be in a set.
Adding an element to a set and testing for membership are both constant time
operations. Bloom filters that have too many elements contained within them are said to be ``too full'', which renders them unusable due to their excessively high FP rate.

Bloom filters work by using a small number of (non-cryptographic) hash functions
that map the set element to a random bit position in the filter. So if you have
five hash functions, there are five positions in the filter for each element,
and this set of positions is characteristic to each element. To 
insert an element into the set, one simply makes all the element's
characteristic positions to one (i.e., ``on''). Thus, to see if an element is in
the set, one simply checks if all the characteristic positions are ``on'': if
any are ``off'' (i.e., not set to 1 but instead still 0) then we are certain
that the element was never inserted. If all are on, however, it may be due to a
coincidence from the other elements that were inserted. Therefore, there are no
false negatives but there are false positives.

When using Bloom filters in the SPV setting, the FP rate is used as a proxy for
the privacy level of the Bloom filter. An SPV client with ample bandwidth may
choose to have a high FP rate. This means that the full node peer from which
data is downloaded to verify a transaction, cannot accurately determine which
transaction the SPV client is interested in verifying. Bloom filters with an
exceedingly high FP rate will download as much data, and have the same privacy
provisions, as Naive SPV\@. This is because such a Bloom filter would match all
addresses and in effect download the entire blockchain.

In contrast, an SPV client with access to a
minimal amount of bandwidth would create an accurate Bloom filter, by setting a
low FP rate. This would mean that the full node peer would know exactly which
transactions the SPV client is interested in verifying, since only a specific set
of data would be downloaded.  As such, Bloom filters represent a trade-off, which
is configurable by the user, between privacy---the precision of the data
returned from a full node peer---and bandwidth.

Once a Bloom filter has been created, it is used in the construction of the corresponding Merkle block. This consists of a block header and a partial Merkle tree, called a Merkle branch. The Merkle branch consists of all the TXIDs for whom set membership passed in the Bloom filter---and which therefore includes false positives---as well as intermediate Merkle tree hashes, including the Merkle tree root.
Altogether, this allows the SPV client to connect these
TXIDs together and to verify the value of the Merkle tree root located in the block header. A Merkle block is followed by a list of transactions which were set in the Bloom filter, and whose TXIDs were already included in the Merkle branch. 

In practice, users tend to use low FP rates when constructing Bloom filters~\cite{jonas}, thus maintaining virtually no privacy. The reason low FP rates are used is because they allow faster synchronization for SPV clients. SPV synchronisation is a process by which a users' wallet is brought up to date with the current state of the blockchain by reflecting in the wallets' balance the final result of the transactions in which the user was involved. Increasing the FP rate leads to longer synchronization times, since more data needs to be downloaded. This has been commonly observed \cite{mike} to have a negative effect on users' satisfaction with SPV client implementations such as BitcoinJ~\cite{btcj}. This occurs because users are more concerned about the overall performance of their wallet application, rather than their privacy.

\subsection{Private Information Retrieval}

Private Information Retrieval (PIR) allows users to query a database, such that
the database learns nothing about the users' query. The trivial solution to this would be for the client to download
the entire database and perform the query offline. Over the years more efficient PIR protocols have emerged which have significantly improved on this upper bound. We provide a brief overview below.

There exist two classes of PIR: Information Theoretic PIR (IT-PIR) and
Computational PIR (C-PIR). IT-PIR protocols involve replicating the database
among a set of servers. The client makes different queries to different servers
and determines the result of their query from the set of server responses. IT-PIR is information theoretically 
secure, provided that the different servers do not collude with each other, up to a certain threshold. C-PIR protocols, in contrast, require only one server to store the database
and query privacy is guaranteed by cryptographic means.

Each class of PIR has its own advantages and disadvantages. The advantages of multi-server IT-PIR are that it generally incurs smaller
communication and computation costs. This is because IT-PIR treats queries as vectors, and
databases as matrices, on which linear algebra operations of vector-by-matrix
multiplication are performed. As such, the size of the database should be square
in order for the communication cost between client and server to be optimal. IT-PIR is also robust to missing or incorrect database server responses.

The disadvantage of IT-PIR is that the threshold of non-colluding
servers needs to be maintained and it is not clear how this requirement can be enforced in practice, particularly because a database server can mount a Sybil attack.

The advantage of single-server C-PIR is that it relies only on a single server,
however, due to this property, C-PIR is not robust since it cannot overcome
missing or incorrect database server responses.

C-PIR is generally computationally slower when compared to IT-PIR\@. This is because greater computation is required in encoding clients' queries and because the server is performing matrix-by-matrix multiplication---in contrast to vector-by-matrix multiplication in some IT-PIR schemes. Another disadvantage of some C-PIR protocols is that they are based on lattice problems whose security provisions are not well understood. Because of this, clients may have doubts about the privacy of their queries~\cite{devet2014best}.

In our implementation we use a hybrid PIR scheme by Devet and Goldberg~\cite{devet2014best} which combines the advantages 
of IT-PIR and C-PIR, while minimising the disadvantages of either one. The underlying IT-PIR protocol in 
this hybrid scheme is by Devet et al.~\cite{devet2012optimally} which leverages Shamir Secret Sharing and Reed-Solomon decoding to create a robust and Byzantine-fault-tolerant IT-PIR protocol which can withstand up to \textit{t} colluding servers (\textit{t} is called the \textit{privacy level}). The C-PIR scheme used is by Aguilar-Melchor and Gaborit~\cite{aguilar2007lattice}.

The hybrid PIR scheme treats the underlying data as elements of the finite field
GF(2\textsuperscript{8}).  This protocol can be found in the Percy++ \cite{p++}
open source library which implements a number of state-of-the-art PIR protocols.

\section{System Overview}
\label{sec:systemoverview}
In this section we describe our system design, its properties and considered adversarial models. We denote the IT-PIR protocol as $\Phi$ and the C-PIR protocol as $\Theta$ in our hybrid scheme. 

\subsection{System Model}
Our system features the following entities:
\begin{itemize}
	\item \textbf{Ledger:} We assume the existence of an immutable ledger, acting as coarse-grained timestamping service. The ledger's purpose is to discern not yet spent electronic coins (commonly referred to as unspent transaction outputs or short UTXO).
	\item \textbf{User:} A user possesses one or more private keys in the
	ledger and intends to validate the transactions towards the accounts
	(e.g., Bitcoin addresses) controlled by the private keys through SPV.
	\item \textbf{PIR server:} The PIR server responds to the PIR queries from the SPV clients and provides the data required to perform the SPV.
\end{itemize}

We assume users can communicate with PIR servers through the underlying network.

\subsection{High-Level Operation}
\begin{figure}[h]
\centering
\includegraphics[width = 0.7\hsize]{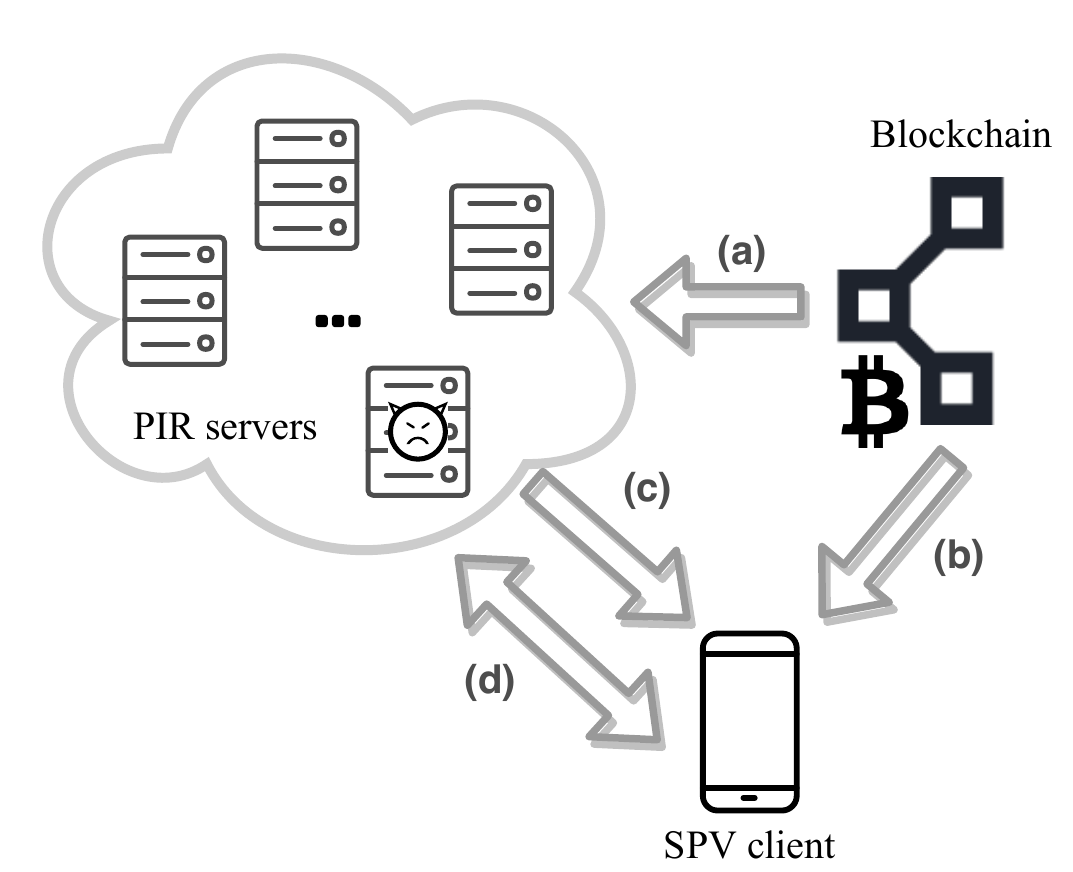}
\caption{High-level operations of our system: (a) PIR servers download the blockchain; (b) User keeps up to date with all available block headers; (c) User fetches the manifest files from the PIR servers; (d) User executes the PIR-SPV protocol.}
\label{fig:highleveloperation}
\end{figure}

In this section, we outline the high-level operations of our system (cf.~Figure~\ref{fig:highleveloperation}).
\begin{itemize}
	\item[(a)] PIR servers download the whole blockchain and construct PIR databases. For each database, the PIR server creates a description file called \textit{manifest file}. 
	\item[(b)] As outlined in Section~\ref{sec:background}, the user collects all
	available block headers from e.g., full node peers. 
	\item[(c)] The user fetches the manifest files from the PIR servers to efficiently query the PIR database afterwards.
	\item[(d)] The user executes the PIR-SPV protocol. PIR queries can be
	run recursively (i.e., in one PIR query, the user encodes and exchanges several query messages before receiving the final PIR response). The user decodes the responses and then performs SPV validation (verifying the partial Merkle tree of the received transactions given the downloaded block headers).
\end{itemize}

\subsection{Main Properties}
Our system provides the following properties:
\begin{itemize}
	\item \textbf{User:} Given our system, users can query their transactions from other blockchain nodes, without disclosing the transactions that they are interested in. We call this property \textit{transaction query privacy}.
	\item \textbf{PIR server:} We assume that different PIR servers create equal databases and manifest files, given the same blockchain source. While serving user requests, the PIR servers do not learn any user-centric transaction data.
\end{itemize}

\subsection{Adversarial Model}
The privacy offered by the hybrid PIR scheme relies on the fact that both schemes $\Phi$ and $\Theta$ preserve the query privacy.
For $\Phi$, we assume that $k$ out of $\ell$ servers respond per PIR query. We then define the privacy level $t$ and the number of Byzantine servers $v$, s.t.\ $v < k - t - 1$. We assume that no more than \textit{t} servers are colluding to discover the contents of a user’s query, which ensures the privacy of $\Phi$. This implies that we assume not all PIR servers are colluding in the case that all servers are responsive and there is no Byzantine server.
For $\Theta$, we assume that malicious PIR servers are computationally bound, which guarantees that disclosing user's interest from a $\Theta$ PIR query is infeasible.


\section{System Details}
\label{sec:systemdetails}
In this section, we detail the architecture of our system. We begin by describing how PIR servers build the databases to facilitate PIR-based queries. Then, we describe the manifest files, which are used by clients to construct PIR-based queries. Finally, we depict our PIR-based SPV protocol.


\subsection{Database Construction}
\subsubsection{Database Content Structure}
\label{sec:databasecontentstructure}
\begin{figure}[h]
\centering
\includegraphics[width = 1\hsize]{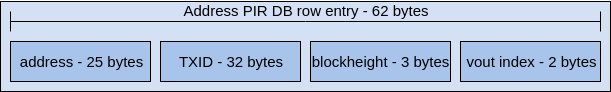}
\caption{Structure of a single Address PIR DB row}
\label{fig:addrpir}
\end{figure}

We split the data required to perform SVP into three distinct components (i.e., the address PIR DB, the Merkle tree PIR DB and the transaction PIR DB), the formats of which are respectively detailed below.
\paragraph{Address PIR DB}
\label{sec:addresspirdb}
In our implementation, we take a user-centric, address-first approach, which means that a user needs to select a Bitcoin address that belongs to them with which they would like to execute our protocol. Figure~\ref{fig:addrpir} illustrates the structure of a single entry in a row in the Address PIR database. A row is made up of one or more of these entries.

An explanation of each field follows:

\begin{itemize}

\item Address: 
This field holds a particular Bitcoin address can be involved in more than one transaction. 
As such, there can be multiple entries with the same address. This field serves as the entry point to our PIR SPV protocol.

\item TXID: This field identifies the transaction whose output address the address field refers to. This field is used to query the Transaction PIR database.

\item Block height: This field holds the block height of the block which includes the transaction the TXID field refers to. 
This field is used to query the Merkle Tree PIR database.

\item Vout Index: This field holds the index value of the output of the transaction referenced by the TXID field, in which the address referenced by the address field was used.  
This field serves more of a convenience than a necessity, by allowing clients to quickly identify the location of their output in the relevant transaction, without having to look through every transaction output.

\end{itemize}

\begin{figure}[h]
\centering
\includegraphics[width = 1\hsize]{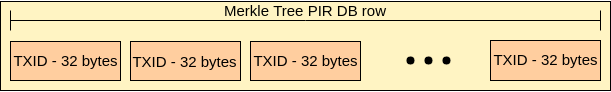}
\caption{Structure of a single Merkle Tree PIR DB row}
\label{fig:mtpir}
\end{figure}

\paragraph{Merkle Tree PIR DB}
\label{sec:merkletreepirdb}
This database contains the TXIDs which are used to calculate the Merkle tree root in the SPV protocol. The structure of a single row in the Merkle Tree PIR database is shown in Figure~\ref{fig:mtpir}, where each TXID is 32 bytes in length.

\begin{figure}[h]
\centering
\includegraphics[width = 1\hsize]{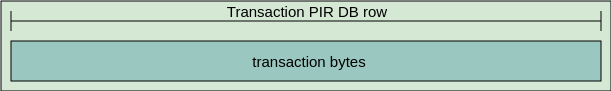}
\caption{Structure of a single Transaction PIR DB row}
\label{fig:rtxpir}
\end{figure}

\paragraph{Transaction PIR DB}
\label{sec:transactionpirdb}
This database contains the hexadecimal format of 
transactions in the UTXO set. Figure~\ref{fig:rtxpir} illustrates the structure of a single row in the Transaction PIR database, where each row consists of transaction bytes.

\subsubsection{Temporal Partitioning of Databases}
We explain our rationale for partitioning the databases across several time periods in this section.

The collected data are split into three distinct time periods, with associated
manifest files: weekly, monthly and all-time. Since a new block is generated
approximately every 10 minutes, weekly data consists of the most-recent $1\,008$
collected blocks, monthly data the preceding $4\,032$ blocks, and all-time data
consisted of all remaining blocks---up to and including the Genesis block.

The temporal partitioning of the collected data means that enormous amounts of data do
not usually need to be searched using PIR\@. This leaks the fact that a client is
interested in, for example, more-recent data whenever the weekly databases are
queried. This can be assumed to be true, however, given the knowledge that an
entity is using Bitcoin; more importantly, simply accessing recent transactions
does not imply one has recent transactions. The important piece
of privacy to maintain is the mapping of a client to the set of addresses that
they control. This is maintained implicitly since the client can only query the
Address PIR DB under PIR.

Clients who are interested in verifying a transaction for a particular range of time simply issue their queries to the corresponding PIR database. This structure allows our static implementation to be easily transformed into one which is periodically updated whenever a new block is generated. This is outlined in Appendix~\ref{app:extendedfuturework}.

The other advantage of this partitioning is that it reduces the bandwidth and latency cost for clients who are interested in verifying more recent transactions.

\subsubsection{Database Dimensions}
In this section, we describe how the height and width of the databases are determined.

Several factors influence the appropriateness of the database dimensions when
using PIR\@.
PIR seeks to minimize the communication cost, which is the bandwidth required to
obtain one row of data from any of our databases. The PIR system that we use has
an optimal communication cost when the ``height'' and ``width'' of the database
are approximately equal. That is, the number of rows is proportional to the size
of each row. This is because requests in PIR have a constant cost per row while
the PIR replies are the size of a single row. As such, we tried to ensure that
our databases were square-like when possible.  Larger sets of data were placed
into rectangular databases to avoid \textit{bandwidth bloat}. Bandwidth bloat is
when individual rows in a database contain a greater proportion of irrelevant
information with respect to the clients query.

It should be noted that larger databases increase the latency of PIR queries, as they are executed over more data which is computationally intensive. In most cases the database size was slightly larger than the set of data being stored in it, resulting in the remaining space being padded with strings of zeros.

\subsubsection{Data Ordering}
Once the dimensions of the databases are determined, the collected data have to be ordered, as described below, before being placed into these databases.

For Address PIR DB data, each entry is lexicographically sorted by the address field. For Merkle Tree PIR DB data, each list of TXIDs is sorted in ascending order according to the associated block height. No ordering is imposed on Transaction PIR DB data.

After the sorting occurred of the respective sets of data, it is then arranged in row-oriented order in the databases. This is where a row is filled from left to right with entries before the next row is filled. This localises the entries, thus reducing the total number of queries that clients need to perform.

\subsection{Database Manifest Files}
In the PIR scheme we utilised, users are required to grasp some preliminary
knowledge of the queried database (e.g., database dimensions and the position of
				   the desired data) before making a request.
Consequently, for each database, a manifest file is generated which provides
sufficient information to bootstrap a PIR query. A manifest file consists of one
or more records, each of which indicates the location of an individual piece of
data. Users request from PIR servers to fetch the manifest files before making
the queries. As manifest files are the global descriptions of the databases, the
request does not reveal any user's interest other than participating in Bitcoin.
Note, in Section~\ref{sec:futurework}, we propose a solution that saves
the trouble of downloading the whole manifest file replaced by iterative PIR
queries to reduce bandwidth cost. We concretely describe the format of the manifest files in Appendix~\ref{app:manifestfilesformats}.


\subsection{PIR-Based SPV Protocol}
\label{sec:pirbasedspvprotocol}


We describe our PIR-based SPV protocol which facilitates
private lightweight Bitcoin clients in Algorithm~\ref{alg:pirspvprotocol}. The client is required to obtain the available block headers independently. Execution of the PIR-SPV protocol is divided into three rounds of queries to three PIR databases in the proper order (cf.~Algorithm~\ref{alg:pirspvprotocol}). Once completing the queries, the user can perform the SPV validation of the selected transaction with the transaction content, TXIDs and the block headers.


\begin{algorithm}[h]
\caption{PIR-SPV protocol}
\label{alg:pirspvprotocol}
\KwData{SPV client $\mathcal{C}$, one or multiple PIR servers $\mathcal{S}$}
\KwResult{$\mathcal{C}$ obtains the necessary data for a SPV privately}
\textbf{Initialization:} $\mathcal{S}$ constructs the PIR databases and associated manifest files; $\mathcal{C}$ downloads the manifest files from $\mathcal{S}$\;
  \nlset{a.1} $\mathcal{C}$ selects an address to fetch a record from the Address PIR DB manifest file and generates the PIR queries based on row indices of the selected record\;
  \nlset{a.2} $\mathcal{S}$ computes the result using the PIR queries on the Address PIR DB\;
  \nlset{a.3} \label{step:ares} $\mathcal{C}$ parses and decodes the result to obtain one or more Address PIR DB entries\;
  \BlankLine
  \nlset{b.1} \label{step:b1} $\mathcal{C}$ uses the value of the block height field of an entry to fetch the corresponding record from the Merkle Tree PIR DB manifest file and generates the PIR queries based on the row indices\;
  \nlset{b.2} $\mathcal{S}$ computes the result using the PIR queries on the Merkle Tree PIR DB\;
  \nlset{b.3} $\mathcal{C}$ parses and decodes the result to obtain the requested list of TXIDs\;
  \BlankLine
  \nlset{c.1} $\mathcal{C}$ uses the value of the TXID field from the same entry
  that was selected in step \ref{step:b1}, to fetch the corresponding record from the Transaction PIR DB manifest file and generates the PIR queries based on the row indices\;
  \nlset{c.2} $\mathcal{S}$ computes the result using the PIR queries on the Transaction PIR DB\;
  \nlset{c.3} $\mathcal{C}$ parses and decodes the result to obtain the requested transaction.
\end{algorithm}

\section{Evaluation}
\label{sec:evaluation}
We collected the necessary data between February and April 2018 from 9 virtual machines (VMs) running the core Bitcoin client. Each VM ran a Ubuntu 16.04 (64-bit) OS, had four $1$ GHz CPUs, $8$ GB of RAM and a $300$ GB hard drive. 
For simplicity, in our implementation, we only considered the UTXO set of P2PKH addresses. Note the Bitcoin Core supports a descriptor scheme\cite{bitcoind83:online} to generally express different types of outputs. We therefore reasonably deduce the feasibility of extending our solution to cover various address types.




Table~\ref{table:dbdimensions} summarises the sizes of the individual components of the system. We use \emph{database width} to denote the number of records stored in a row. For example, the all-time data for the Address PIR DB  states that a single row contains $906$ entries, each of which is $62$ bytes in size (cf.~Section~\ref{sec:systemdetails}).
Because the data fields vary across the different databases, the exact number of bytes that a database row consumes therefore varies. We detail the process of determining the dimensions in  Appendix~\ref{appx:dimensiondetermination}.


\begin{table*}[h!]
\centering
\caption{The sizes and dimensions of the generated databases and their corresponding manifest files}
\label{table:dbdimensions}
\begin{tabular}{lllll}
\toprule
\multicolumn{2}{l}{} &
\multicolumn{1}{c}{\begin{tabular}[c]{@{}c@{}}All-Time\\ blocks 1 to
	508462\end{tabular}} &
	\multicolumn{1}{c}{\begin{tabular}[c]{@{}c@{}}Monthly\\ blocks 508463
		to 512494\end{tabular}} &
		\multicolumn{1}{c}{\begin{tabular}[c]{@{}c@{}}Weekly\\ blocks
			512495 to 513502\end{tabular}} \\ \midrule
    Address PIR DBs & \begin{tabular}[c]{@{}l@{}}\# entries per row (width)\\ \#
    rows (height)\\ Total size of DB\\ Size of manifest file\end{tabular} &
    \begin{tabular}[c]{@{}l@{}}906\\ 56172\\ 3.16 GB\\ 65.99 MB\end{tabular} &
    \begin{tabular}[c]{@{}l@{}}214\\ 13268\\ 176.04 MB\\ 175.33 MB\end{tabular}
    & \begin{tabular}[c]{@{}l@{}}124\\ 7688\\ 59.11 MB\\ 66.97 MB\end{tabular}
    \\ \midrule
Merkle Tree PIR DBs & \begin{tabular}[c]{@{}l@{}}\# entries per row (width)\\ \#
rows (height)\\ Total size of DB\\ Size of manifest file\end{tabular} &
\begin{tabular}[c]{@{}l@{}}821\\ 394080\\ 10.18 GB\\ 40.90 MB\end{tabular} &
\begin{tabular}[c]{@{}l@{}}1196\\ 38272\\ 1.46 GB\\ 0.32 MB\end{tabular} &
\begin{tabular}[c]{@{}l@{}}1184\\ 37888\\ 1.44 GB\\ 78.62 KB\end{tabular} \\
	\midrule
Transaction PIR DBs & \begin{tabular}[c]{@{}l@{}}Length of row in bytes
(width)\\ \# rows (height)\\ Total size of DB\\ Size of manifest
file\end{tabular} & \begin{tabular}[c]{@{}l@{}}758\\ 20942782\\ 15.87 GB\\ 3.03
GB\end{tabular} & \begin{tabular}[c]{@{}l@{}}848\\ 1537424\\ 1.30 GB\\ 218.68
MB\end{tabular} & \begin{tabular}[c]{@{}l@{}}876\\ 512460\\ 448.91 MB\\ 72.45
MB\end{tabular} \\ \bottomrule
\end{tabular}
\end{table*} 
\subsection{Benchmark Methodology}
\label{sec:benchmarkmethodology}
We executed our PIR-based SPV protocol and compared its bandwidth cost with the Bloom-filter-based BIP-37 SPV protocol and with the fully private Naive SPV protocol. Since we use Devet et al.'s \cite{devet2014best} hybrid PIR scheme, the client can query one or more PIR servers when executing our protocol. The experiments were executed locally on a single machine, running one or more PIR server instances. This machine had a Ubuntu 16.04 (64-bit) OS, Intel Core i7 3.4 GHz CPU, 16 GB of RAM and a 512 GB hard drive. To ensure a fair and accurate comparison, we did not include the cost of downloading the block headers, since this is a common element of all three protocols. We did not include the bandwidth cost of client-side queries for BIP-37 and Naive SPV since these are contained in peer-to-peer layer messages. We did, however, include the bandwidth cost of client-side queries for PIR SPV since that is a fundamental element of performing PIR\@. The other steps that we followed when collecting our data were:

\begin{itemize}
\item \textbf{PIR SPV}: For every TXID located in each entry in the Address PIR
DBs, we executed the PIR protocol as outlined in Section~\ref{sec:systemdetails} and used the total length of the returned rows, from one or more PIR servers, as part of the bandwidth cost. The downloading of manifest files is not included in our calculation of bandwidth cost because in Section~\ref{sec:futurework} we present a feasible solution which removes the need for clients to download them. Instead, clients perform interpolation search using iterative PIR queries to retrieve a particular record, which we imagine will have a small bandwidth cost.
 
Two sets of data were collected for PIR SPV\@. The first set of data used only one
PIR server to simulate the case of single-server C-PIR being used by the hybrid
PIR schema. The second set of data used three PIR servers to simulate the case
of multi-server IT-PIR being used by the hybrid PIR schema. The bandwidth cost
of using additional servers is linear, assuming that no Byzantine PIR servers
exist. It should be noted that under C-PIR, a recursive depth of size $1$ was
used, as suggested by Aguilar-Melchor and Gaborit \cite{aguilar2007lattice}.

\item \textbf{BIP-37 SPV}: For every TXID located in each entry in the Address
PIR DBs, a Bloom filter was constructed and then used to send a Bitcoin peer-to-peer layer message to a full node hosted on one of our VM’s. The length of the reply, which contained the Merkle block and the associated transactions, was measured and used as the bandwidth cost. It should be noted that this reply also contained a network message header which was $24$ bytes in length \cite{btcmsgprotoc}. This value was deducted from the final bandwidth cost calculations since this does not exist in the other two protocols. 

\item \textbf{Naive SPV}: For every TXID located in each entry in the Address
PIR DBs, we measured the total amount of bandwidth that had to be used to download the blocks between the Genesis block and the block in which the transaction of interest was included. We used a single full node hosted on one of our VMs.

\end{itemize}

\subsection{Results and Analysis}

\begin{figure*}[htb!]
    \centering 
\begin{subfigure}{1\textwidth}
  \includegraphics[width=\linewidth]{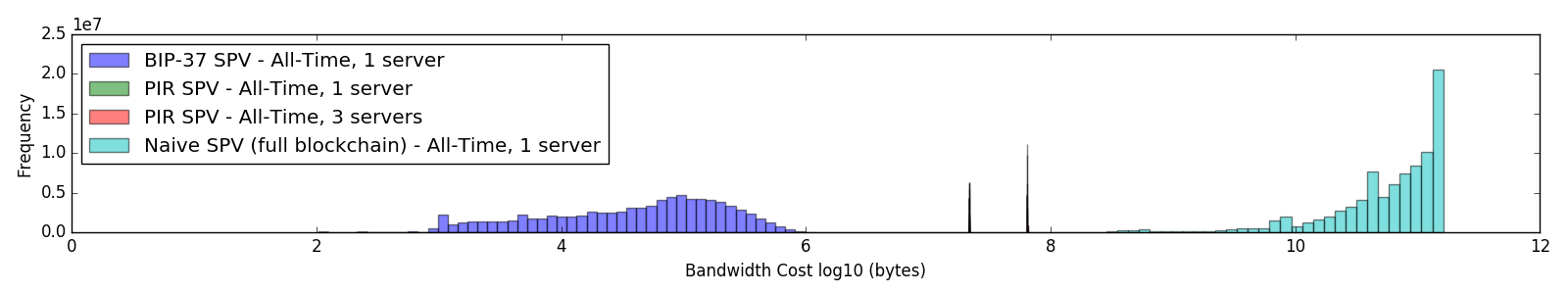}
  \caption{All-Time}
  \label{fig:1}
\end{subfigure}\hfil 
\begin{subfigure}{1\textwidth}
  \includegraphics[width=\linewidth]{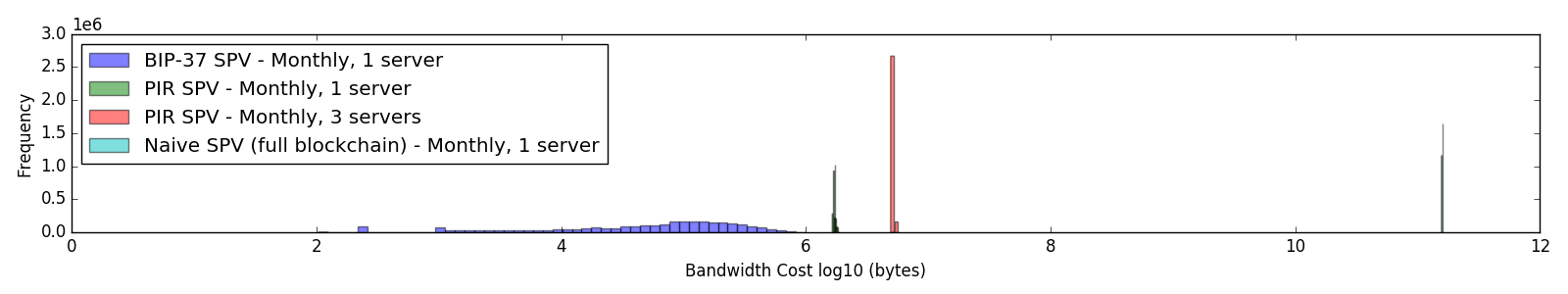}
  \caption{Monthly}
  \label{fig:2}
\end{subfigure}\hfil 
\begin{subfigure}{1\textwidth}
  \includegraphics[width=\linewidth]{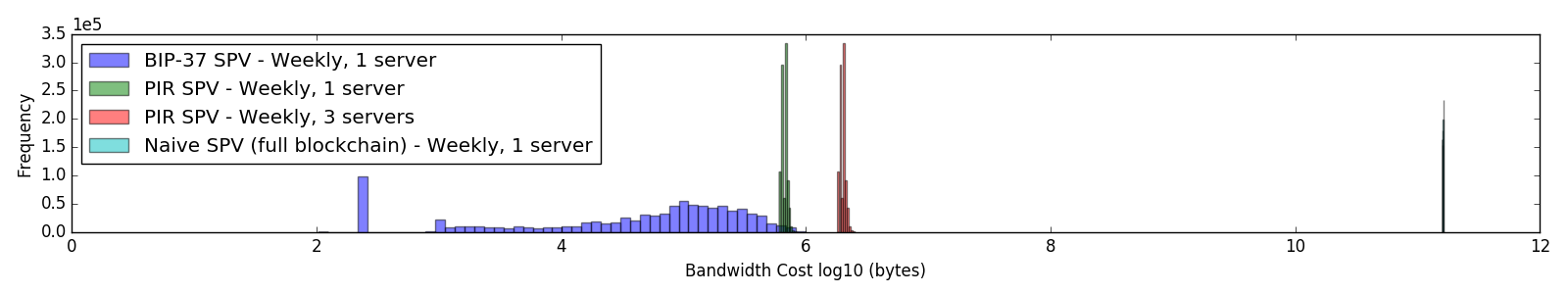}
  \caption{Weekly}
  \label{fig:3}
\end{subfigure}

\captionsetup{width=1\textwidth}

\caption{Histograms showing the bandwidth cost of verifying a single TXID by BIP-37 SPV, single- and multi-server PIR SPV and Naive SPV, for the three time periods.}

\label{fig:bandwidthhist}
\end{figure*}

The results are recorded in Figure~\ref{fig:bandwidthhist}, which shows three histograms displaying the bandwidth cost required to verify a single transaction for each of the different SPV protocols, across all-time, monthly and weekly data. These results are summarised in Table~\ref{table:bandwidthcost}, which shows the expectation and standard deviation of the collected data.

\begin{table*}[h!]
\centering
\caption{Summary of Figure~\ref{fig:bandwidthhist}, showing the expectation and standard deviation of the bandwidth cost for the different protocols, for the three time periods.
}
\label{table:bandwidthcost}
\begin{tabular}{llllll}
\toprule
\multicolumn{2}{l}{} & \multicolumn{1}{c}{BIP-37} & \multicolumn{1}{c}{PIR: 1
	server} & \multicolumn{1}{c}{PIR: 3 servers} & \multicolumn{1}{c}{Naive}
	\\ \midrule

All-Time & \begin{tabular}[c]{@{}l@{}}expectation\\ std.~dev.\end{tabular} &
\begin{tabular}[c]{@{}l@{}}102.80 KB\\ 130.25 KB \end{tabular} &
\begin{tabular}[c]{@{}l@{}}21.54 MB\\ 35.43 KB\end{tabular} &
\begin{tabular}[c]{@{}l@{}}64.61 MB\\ 106.29 KB\end{tabular} &
\begin{tabular}[c]{@{}l@{}}80.27 GB\\ 50.24 GB\end{tabular}\\ \midrule

 
Monthly & \begin{tabular}[c]{@{}l@{}}expectation\\ std.~dev.\end{tabular} &
\begin{tabular}[c]{@{}l@{}}132.43 KB\\ 149.14 KB \end{tabular} &
\begin{tabular}[c]{@{}l@{}}1.70 MB\\ 34.00 KB\end{tabular} &
\begin{tabular}[c]{@{}l@{}}5.11 MB\\ 102.01 KB\end{tabular} &
\begin{tabular}[c]{@{}l@{}}159.19 GB\\ 1.12 GB\end{tabular}\\  \midrule

 
Weekly & \begin{tabular}[c]{@{}l@{}}expectation\\ std.~dev.\end{tabular} & \begin{tabular}[c]{@{}l@{}}128.52 KB\\ 155.19 KB\end{tabular} & \begin{tabular}[c]{@{}l@{}}666.07 KB\\ 34.00 KB\end{tabular} & \begin{tabular}[c]{@{}l@{}}2.00 MB\\ 102.66 KB\end{tabular} & \begin{tabular}[c]{@{}l@{}}161.47 GB\\ 292.08 MB\end{tabular}\\
\bottomrule
\end{tabular}
\end{table*}

\begin{figure*}[htb!]
    \centering 
\begin{subfigure}{0.33\textwidth}
  \includegraphics[width=\linewidth]{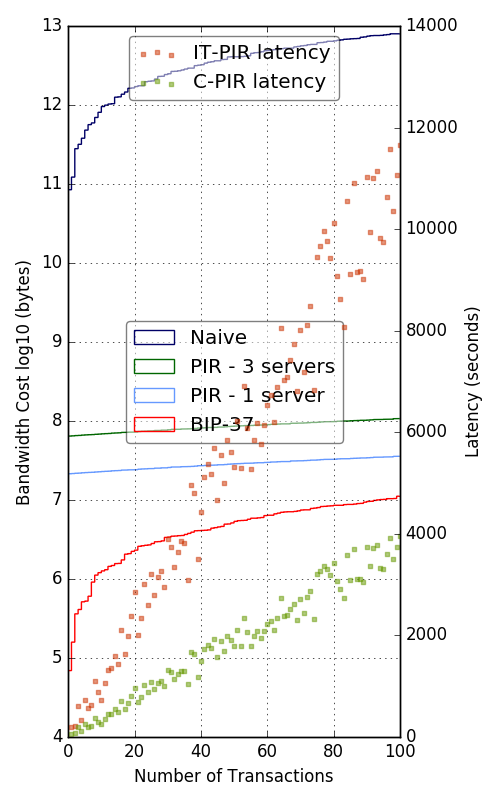}
  \caption{All-Time}
  \label{fig:1}
\end{subfigure}\hfil 
\begin{subfigure}{0.33\textwidth}
  \includegraphics[width=\linewidth]{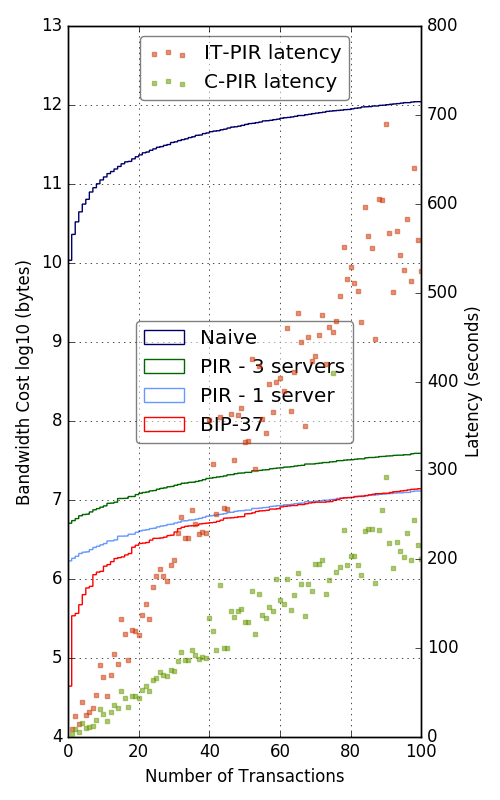}
  \caption{Monthly}
  \label{fig:2}
\end{subfigure}\hfil 
\begin{subfigure}{0.33\textwidth}
  \includegraphics[width=\linewidth]{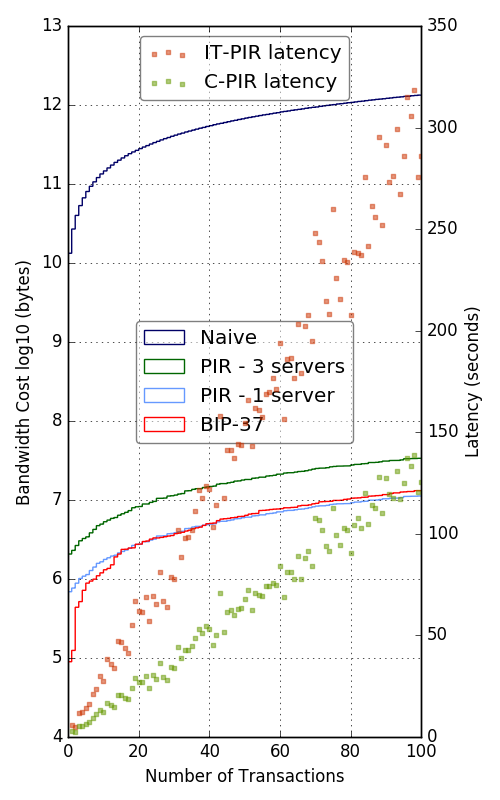}
  \caption{Weekly}
  \label{fig:3}
\end{subfigure}

\captionsetup{width=1.0\textwidth}
\centering 
\caption{Cumulative distribution of bandwidth cost for the verification of transactions under BIP-37 SPV, single- and multi-server PIR SPV and Naive SPV, across the three time periods. The latency of single- and multi-server PIR SPV is also displayed.}

\label{fig:bandwidthcdf}
\end{figure*}

Figure~\ref{fig:bandwidthcdf} shows the cumulative bandwidth cost required to
verify up to 100 transactions by the different SPV protocols, across the three
time periods. The corresponding latency cost of executing our PIR SPV protocol
in the single- and multi-server setting is also shown. A small sample of these results is shown in Table~\ref{table:cdfsummary}.
In Figure~\ref{fig:bandwidthcdf}, to calculate the bandwidth cost of verifying
one or more transactions through SPV, steps similar to those described in
Section~\ref{sec:benchmarkmethodology} were followed with some adjustments. For
example, to calculate the total bandwidth cost of verifying 7 transactions, a
random entry from the Address PIR DB was picked 7 times, and its TXID used to
perform Naive, BIP-37 and PIR-based SPV\@. The bandwidth cost for each TXID was
calculated as described in Section~\ref{sec:benchmarkmethodology} and then the
results of the 7 calculations were totaled. We did this five times and then an
average was taken. This average is used as the final result.

It should be noted that when calculating the bandwidth cost of Bloom-filter-based SPV, a fresh filter was used on each TXID, rather than a single filter encoding the set of 7 TXIDs. This is because a Bloom filter gets full very quickly when it is used to match a large number of TXIDs. A full Bloom filter is one where most of the entries are set to ``on''. Using a fresh filter for each TXID would mean that we would obtain a more accurate bandwidth cost.

The latency was measured by executing the hybrid PIR protocol from Percy++ \cite{p++} on the average number of rows needed to verify a set of TXIDs, for example 7, for each of the Address, Merkle Tree and Transaction PIR databases. Figure~\ref{fig:bandwidthcdf} shows the latency of a single server, for both the C-PIR and IT-PIR cases, because even though 3 servers were used to facilitate IT-PIR, the query was sent in parallel to each server rather than sequentially.

In the weekly and monthly case, Figure~\ref{fig:bandwidthcdf} shows that single-server PIR SPV has a similar bandwidth cost to BIP-37 SPV when verifying between 20 and 100 transactions. In particular, Table \ref{table:cdfsummary} shows that if a client is interested in performing single-sever PIR SPV on 100 transactions which occurred in the past week, it will take 
approximately 2 minutes for this query to be executed, with a bandwidth cost of 11.18 MB. In contrast, BIP-37 SPV will require 12.85 MB in the same scenario.

\begin{table*}[h!]
\centering
\caption{A small sample of results from Figure~\ref{fig:bandwidthcdf}, showing the bandwidth and latency cost for a particular number of transactions under different SPV protocols, across the three time periods.}
\label{table:cdfsummary}
\begin{tabular}{llllllll}
\toprule
\multicolumn{2}{c}{Number of Transactions} & \multicolumn{1}{c}{BIP-37} &
\multicolumn{1}{c}{PIR: 1 server} & \multicolumn{1}{c}{PIR: 3 servers} &
\multicolumn{1}{c}{Naive} & \multicolumn{1}{c}{C-PIR latency} &
\multicolumn{1}{c}{IT-PIR latency} \\ \midrule
1 & \begin{tabular}[c]{@{}l@{}}All-Time\\ Monthly\\Weekly\end{tabular}&
\begin{tabular}[c]{@{}l@{}}69.32 KB\\ 44.08 KB\\89.78 KB\end{tabular}&
\begin{tabular}[c]{@{}l@{}}21.51 MB\\ 1.70 MB\\691.04 KB\end{tabular}&
\begin{tabular}[c]{@{}l@{}}64.53 MB\\ 5.09 MB\\2.07 MB\end{tabular}&
\begin{tabular}[c]{@{}l@{}}84.85 GB\\ 10.86 GB\\13.32 GB\end{tabular}&
\begin{tabular}[c]{@{}l@{}}68.44 s\\ 3.89 s\\2.84 s\end{tabular}&
\begin{tabular}[c]{@{}l@{}}203.52 s\\ 9.39 s\\6.02 s\end{tabular}
\\ \midrule


20 & \begin{tabular}[c]{@{}l@{}}All-Time\\ Monthly\\Weekly\end{tabular}&
\begin{tabular}[c]{@{}l@{}}2.82 MB\\ 2.87 MB\\2.76 MB\end{tabular}&
\begin{tabular}[c]{@{}l@{}}23.41 MB\\ 4.07 MB\\2.77 MB\end{tabular}&
\begin{tabular}[c]{@{}l@{}}70.23 MB\\ 12.20 MB\\8.31 MB\end{tabular}&
\begin{tabular}[c]{@{}l@{}}1.72 TB\\ 234.94 GB\\282.31 GB\end{tabular}&
\begin{tabular}[c]{@{}l@{}}680.54 s\\ 53.07 s\\26.85 s\end{tabular}&
\begin{tabular}[c]{@{}l@{}}2008.16 s\\ 137.28 s\\61.78 s\end{tabular}
\\ \midrule


40 & \begin{tabular}[c]{@{}l@{}}All-Time\\ Monthly\\Weekly\end{tabular}&
\begin{tabular}[c]{@{}l@{}}3.80 MB\\ 5.18 MB\\5.50 MB\end{tabular}&
\begin{tabular}[c]{@{}l@{}}25.46 MB\\ 6.32 MB\\4.89 MB\end{tabular}&
\begin{tabular}[c]{@{}l@{}}76.37 MB\\ 18.97 MB\\14.68 MB\end{tabular}&
\begin{tabular}[c]{@{}l@{}}3.16 TB\\ 461.65 GB\\549.78 GB\end{tabular}&
\begin{tabular}[c]{@{}l@{}}1736.52 s\\ 119.43 s\\45.31 s\end{tabular}&
\begin{tabular}[c]{@{}l@{}}5124.89 s\\ 307.34 s\\103.20 s\end{tabular}
\\ \midrule


60 & \begin{tabular}[c]{@{}l@{}}All-Time\\ Monthly\\Weekly\end{tabular}&
\begin{tabular}[c]{@{}l@{}}6.38 MB\\ 8.20 MB\\8.58 MB\end{tabular}&
\begin{tabular}[c]{@{}l@{}}27.44 MB\\ 8.53 MB\\7.09 MB\end{tabular}&
\begin{tabular}[c]{@{}l@{}}82.32 MB\\ 25.60 MB\\21.27 MB\end{tabular}&
\begin{tabular}[c]{@{}l@{}}4.91 TB\\ 682.80 GB\\821.44 GB\end{tabular}&
\begin{tabular}[c]{@{}l@{}}2281.91 s\\ 149.21 s\\68.89 s\end{tabular}&
\begin{tabular}[c]{@{}l@{}}6732.03 s\\ 389.73 s\\156.38 s\end{tabular}
\\ \midrule


80 & \begin{tabular}[c]{@{}l@{}}All-Time\\ Monthly\\Weekly\end{tabular}&
\begin{tabular}[c]{@{}l@{}}8.31 MB\\ 10.80 MB\\9.12 MB\end{tabular}&
\begin{tabular}[c]{@{}l@{}}29.61 MB\\ 10.81 MB\\9.26 MB\end{tabular}&
\begin{tabular}[c]{@{}l@{}}88.83 MB\\ 32.44 MB\\27.79 MB\end{tabular}&
\begin{tabular}[c]{@{}l@{}}6.44 TB\\ 898.70 GB\\1.09 TB\end{tabular}&
\begin{tabular}[c]{@{}l@{}}3079.08 s\\ 203.77 s\\104.49 s\end{tabular}&
\begin{tabular}[c]{@{}l@{}}9083.17 s\\ 511.47 s\\238.56 s\end{tabular}
\\ \midrule


100 & \begin{tabular}[c]{@{}l@{}}All-Time\\ Monthly\\Weekly\end{tabular}&
\begin{tabular}[c]{@{}l@{}}10.09 MB\\ 13.50 MB\\12.85 MB\end{tabular}&
\begin{tabular}[c]{@{}l@{}}31.50 MB\\ 12.85 MB\\11.18 MB\end{tabular}&
\begin{tabular}[c]{@{}l@{}}94.49 MB\\ 38.56 MB\\33.54 MB\end{tabular}&
\begin{tabular}[c]{@{}l@{}}8.03 TB\\ 1.11 TB\\1.34 TB\end{tabular}&
\begin{tabular}[c]{@{}l@{}}3948.97 s\\ 200.78 s\\125.31 s\end{tabular}&
\begin{tabular}[c]{@{}l@{}}11650.12 s\\ 523.98 s\\286.26 s\end{tabular}
\\ 

\bottomrule
\end{tabular}
\end{table*}

Clients who are interested in a smaller number of transactions for the same time period, such as 20, would incur a bandwidth cost of 8.31 MB in the multi-server setting, with a latency of approximately 62 seconds. In the monthly case, a similar query would incur a bandwidth cost of 12.20 MB with a latency of approximately 2.3 minutes, while in the all-time case this would increase to 70.23 MB and approximately 34 minutes respectively. In the same scenario, BIP-37 SPV will require only 2.76, 2.87 and 2.82 MB respectively.

We expect that clients will query monthly and weekly data most often and that all-time data will only be queried when the client is synchronising with the Bitcoin blockchain for the first time, and when the client has been disconnected from the Bitcoin blockchain for more than a month. We imagine that these scenarios will happen infrequently, and as such the user experience will not be adversely affected by the large latency of querying all-time data.

\section{Future Work}
\label{sec:futurework}
In this section, we mainly discuss the future work on the \emph{manifest file trie} scheme, which attempts to spare the manifest file downloading bandwidth. We further discuss other extensions (i.e.\ database partitioning, dynamic protocol and integration with bitcoin) in Appendix~\ref{app:extendedfuturework}. 

Since in some cases most of the PIR database entries hold distinct identities (e.g.\ the weekly address PIR database), which requires a single record in the manifest file for each entry in the database, manifest files have a nearly equivalent or even larger size compared with the database (cf.\ Table~\ref{table:dbdimensions}). This may incur a poor performance in bandwidth because the client is required to download the whole manifest file.

Manifest files are currently structured as a list of entries which need to be sent to the client in order for the client to be able to query the corresponding databases. Sending the client an updated copy of these manifest files whenever a server-side change occurs (which would be every 10 minutes) is not bandwidth efficient. In order to mitigate this problem, we propose transforming the simple manifest files into a format suitable for PIR queries. These PIR compatible manifest files would then be stored on the PIR server and clients, in order to extract a particular record, would perform interpolation search on these manifest files, under PIR.

With this approach we remove the excessive bandwidth cost of sending full manifest files to the client, and ensure that the record of interest is extracted privately. The addition of PIR-based manifest files requires a slight update to our protocol, whereby the client would first need to perform interpolation search under PIR in order to select each record from each of the manifest files, before using that record in our protocol as outlined in Section~\ref{sec:systemdetails}.

\section{Related Work}
\label{sec:relatedwork}
There exists a limited body of work which focuses on solving the problem associated with the privacy provisions of Bloom filters for the SPV protocol.

In 2017, Kanemura et al.~\cite{kanemura2017design} proposed a $\gamma$-deniability enabled Bloom filter to be used instead of the current implementation as specified in BIP-37. $\gamma$-deniability is a privacy metric that shows how many true positives are hidden by false positives. In this case, for the number of Bitcoin addresses or TXIDs set in a Bloom filter. Hence, a higher $\gamma$-deniability value would indicate a greater level of privacy since more true positives would be obscured by false positives. 
There exists an alternative proposal detailed in BIP-157~\cite{bip157} which describes client-side block filtering. This approach is the inverse of BIP-37. Rather than having clients send a filter to a full node, full node peers generate deterministic filters on each block of transactions that they store and send these to the client. The client is then able to see if this deterministic filter matches any data it is interested in. In the case that there is a positive match, the client requests for the relevant full block of data to be downloaded. Compared to BIP-37, the privacy provision of this proposal is greatly improved. This is because full node peers are only able to determine which full block the client is interested in, compared to a Bloom filter, which identifies individual addresses and transactions a client \emph{may} be interested in. 

Compared to solutions based on probabilistic data structures, the only information leakage in our implementation is that of time since clients can choose which time period they would like to query. It can be easily assumed that all clients would like to query the most recent data found in the weekly databases, and the monthly and all-time set's of data are too large for an adversary to gain any meaningful information. In addition, our protocol uses less bandwidth since the PIR queries are precise. Since each block of transactions has a maximum size of 1 MB in Bitcoin, client-side block filtering will require the client to download at least 1 MB of data in the worst case. 
If a client would like to verify 20 transactions which occurred in the past week, they would incur a bandwidth cost of 8.31 MB by using our protocol, in the worst case. If these transactions are located in individual blocks, then the bandwidth cost of BIP-157 would be 20 MB in the worst case.

Matetic et al.~\cite{matetic2018bite} and W{\"u}st et al.~\cite{wust2019zlite} proposed approaches to protect the privacy of lightweight clients, which however requires a trusted execution environment.

Other work has focused on increasing the privacy of Bitcoin transactions. Dandelion~\cite{dandy}, which is specified in BIP-156, proposes privacy enhanced routing for transactions, but is yet to be integrated into the core Bitcoin client. Currently, when a transaction is submitted into the Bitcoin peer-to-peer network, an adversary can easily link an IP address to the node which broadcasts that particular transaction. Dandelion mitigates this class of attacks by sending transactions over a randomly selected path in the Bitcoin peer-to-peer network before broadcasting that transaction to other peers.

There are some other protocols attempting to mitigate the privacy issues (e.g., Monero~\cite{monero}, Zcash~\cite{zcash}, Solidus~\cite{abraham2016solidus, cecchetti2017solidus} and Cryptonote~\cite{van2013cryptonote}). 
These techniques ensure that even though the blockchain of these cryptocurrencies is public, information regarding the source, amount and destination of a transaction is obfuscated and cannot be deduced. For further analysis of their effectiveness, we refer the reader to~\cite{moser2018empirical, kumar2017traceability, kappos2018empirical}.


\section{Conclusion}
\label{sec:conclusion}
Bitcoin's lack of privacy with the use of Bloom filters for SPV clients has been a known issue for many years. Our work solves this problem by applying PIR to the SPV protocol and we develop an implementation to validate our design. We used this software to measure the bandwidth cost and latency of our protocol and found that its bandwidth cost was similar to that of BIP-37 SPV in the single-server setting of the hybrid PIR schema, and comparable in the multi-server setting. We also proposed some improvements to our protocol, one of which removed the bandwidth cost of downloading large manifest files. We believe we are the first to detail a fully private practical protocol for SPV in Bitcoin.

\bibliographystyle{IEEEtran}
\bibliography{bibliography}
%


\appendix
\section{Appendix}
\subsection{Manifest Files Formats}
\label{app:manifestfilesformats}
In this section, we describe the formats for the three databases' manifest files.
\subsubsection{Address PIR DB Manifest File}

Each record in this manifest file consists of a mapping of an address to the location of the corresponding row entry(ies) in the Address PIR DB. Figure~\ref{fig:mfa} illustrates the structure of this record.

\begin{figure}[h]
\centering
\begin{subfigure}{0.30\textwidth}
\{... \linebreak 
\hspace*{0.2in}(\romannumeral1) \hfill``address'' : [\hfill\hfill\hfill\hfill\linebreak 
\hspace*{0.2in}(\romannumeral2) ``row\_index\_start'',\hspace*{0.2in}\linebreak 
\hspace*{0.2in}(\romannumeral3) ``row\_index\_end'',\hspace*{0.2in}\linebreak 
\hspace*{0.2in}(\romannumeral4) \hspace*{5pt} ``column\_index\_start'',\linebreak 
\hspace*{0.2in}(\romannumeral5) \hspace*{0.2in} ``column\_index\_end'',\linebreak 
\hspace*{0.05in} ]\hspace*{1.4in}\linebreak 
...\}
\end{subfigure}\hfil 

\caption{Example record in Address PIR DB manifest file}
\label{fig:mfa}

\end{figure}

\begin{enumerate}
\item[(\romannumeral1)] The address contained in the address field of the entry referenced by this record. 
\item[(\romannumeral2)] The row index where the first entry is located.
\item[(\romannumeral3)] The row index where the last entry is located.
\item[(\romannumeral4)] The column index where the first entry is located.
\item[(\romannumeral5)] The column index where the last entry is located.
\end{enumerate}

\subsubsection{Merkle Tree PIR DB Manifest File}

Each record in this manifest file consists of a mapping of a block height to the location of 
the corresponding list of TXIDs in the Merkle Tree PIR DB. Figure~\ref{fig:mfm} illustrates the structure of this record.

\begin{figure}[h!]
\centering
\begin{subfigure}{0.30\textwidth}
\{... \linebreak 
\hspace*{0.2in}(\romannumeral1) \hfill``block-height'' : [\hfill\hfill\hfill\hfill\linebreak 
\hspace*{0.2in}(\romannumeral2) ``row\_index\_start'',\hspace*{0.2in}\linebreak 
\hspace*{0.2in}(\romannumeral3) ``row\_index\_end'',\hspace*{0.2in}\linebreak 
\hspace*{0.2in}(\romannumeral4) \hspace*{5pt} ``column\_index\_start'',\linebreak 
\hspace*{0.2in}(\romannumeral5) \hspace*{0.2in} ``column\_index\_end'',\linebreak 
\hspace*{0.05in} ]\hspace*{1.4in}\linebreak 
...\}
\end{subfigure}\hfil 
\caption{Example record in Merkle Tree PIR DB manifest file}
\label{fig:mfm}
\end{figure}

\begin{enumerate}
\item[(\romannumeral1)] The block height of the TXID’s referenced by this record. 
\item[(\romannumeral2)] The row index where the first TXID is located.
\item[(\romannumeral3)] The row index where the last TXID is located.
\item[(\romannumeral4)] The column index where the first TXID is located.
\item[(\romannumeral5)] The column index where the last TXID is located.
\end{enumerate}

\subsubsection{Transaction PIR DB Manifest File}

Each record in this manifest file consists of a mapping of a TXID to the location of the corresponding transaction in the Transaction PIR DB. Figure~\ref{fig:mft} illustrates the
structure of this record.

\begin{figure}[h!]
\centering
\begin{subfigure}{0.30\textwidth}
\{... \linebreak 
\hspace*{0.2in}(\romannumeral1) \hfill``txid'' : [\hfill\hfill\hfill\hfill\linebreak 
\hspace*{0.2in}(\romannumeral2) ``row\_index\_start'',\hspace*{0.2in}\linebreak 
\hspace*{0.2in}(\romannumeral3) ``row\_index\_end'',\hspace*{0.2in}\linebreak 
\hspace*{0.2in}(\romannumeral4) \hspace*{5pt} ``column\_index\_start'',\linebreak 
\hspace*{0.2in}(\romannumeral5) \hspace*{0.2in} ``column\_index\_end'',\linebreak 
\hspace*{0.05in} ]\hspace*{1.4in}\linebreak 
...\}
\end{subfigure}\hfil 
\caption{Example record in Transaction PIR DB manifest file}
\label{fig:mft}
\end{figure}

\begin{enumerate}
\item[(\romannumeral1)] The TXID of the transaction referenced by this record. 
\item[(\romannumeral2)] The row index where the first byte of the transaction is located.
\item[(\romannumeral3)] The row index where the last byte of the transaction is located. 
\item[(\romannumeral4)] The column index where the first byte of the transaction is located.
\item[(\romannumeral5)] The column index where the last byte of the transaction is located.
\end{enumerate}

\subsection{Database Dimensions Determination}
\label{appx:dimensiondetermination}
Below we detail the methodology we followed for determining the dimensions of the databases.

\paragraph{Address PIR DB Dimensions}
The size of the all-time, monthly and weekly data was measured and the height and width values were chosen such that the databases were square in shape, to minimise the communication cost.

\begin{figure*}[h]
    \centering
    \begin{minipage}{0.49\textwidth}
        \centering
        \includegraphics[width=0.9\textwidth]{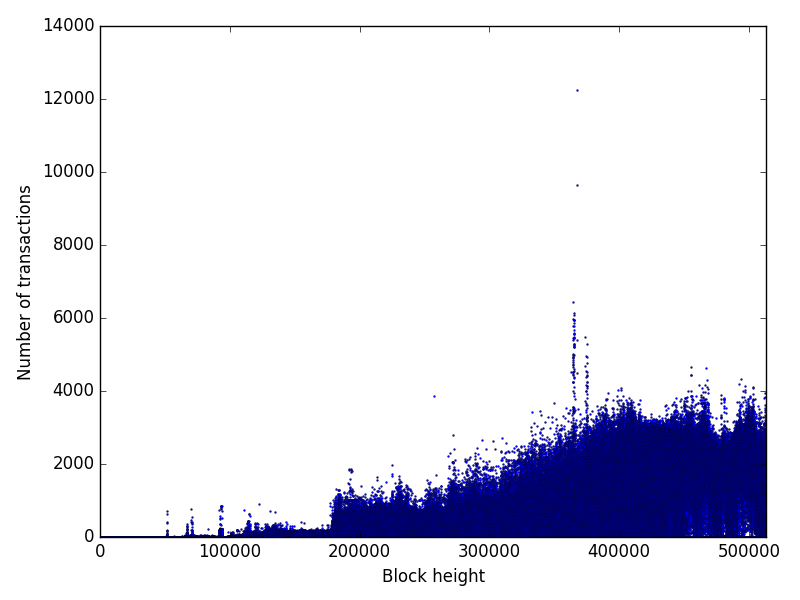}
        \caption{Number of TXID's per block}
    \label{fig:mktpirsize}
    \end{minipage}\hfill
    \begin{minipage}{0.49\textwidth}
        \centering
        \includegraphics[width=0.9\textwidth]{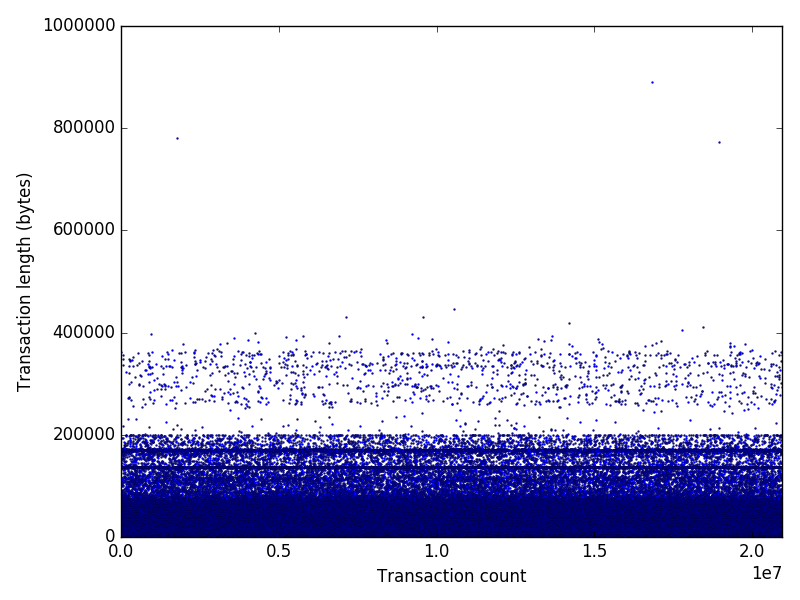}
        \caption{Length of transactions over time}
    \label{fig:txlen}
    \end{minipage}
\end{figure*}


\paragraph{Merkle Tree \& Transaction PIR DB Dimensions}

To determine the width of the rows for these two database types, two graphs were plotted. Figure~\ref{fig:mktpirsize} shows the number of transactions per block and Figure~\ref{fig:txlen} shows how the length of transactions changes over time. For both database types, for the three time periods, the width was taken as the expected value of the respective data sets. The statistical measure of expectation was used for the width because this would reduce the number of times a query would have to be performed, given that in expectation, the row result given to the client would contain the complete set of data required. However, the width of the Merkle Tree PIR DB for the set of all-time data was taken as the running average instead, since Figure~\ref{fig:mktpirsize} shows an extremely skewed distribution. Monthly and weekly Merkle Tree DBs were square, while the rest, namely the all-time Merkle Tree DB and all Transaction DBs, were rectangular.

\subsection{Extended Future Work}\label{app:extendedfuturework}
\subsubsection{Database Partitioning}
Since all-time databases are notably large, partitioning them would reduce the bandwidth and latency cost of queries, since they will be executed over a smaller data set. Database partitioning would require for two new fields, ``pir\_db\_start'' and ``pir\_db\_end'', to be added to the respective manifest files, to track the databases in which a particular entry is included. This would require the PIR based SPV protocol to be slightly adjusted, with the client selecting both the database and row indices when constructing queries. However, database partitioning would expose the client to a statistical privacy attack. Here, a malicious PIR server can monitor for a pattern of querying across multiple databases by the same client and attempt to deduce the user's interests. As such, the partitioning should not be excessive. As a guideline, each partitioned all-time sub-database should not be smaller than the size of the corresponding database for the monthly set of data.


\subsubsection{Dynamic Protocol}
Currently our proof-of-concept implementation is static, which means that it does not reflect the latest state of the Bitcoin blockchain. A new block of transactions is mined and sent out into the Bitcoin peer-to-peer network approximately once every 10 minutes. 

In order to create a dynamic implementation of our protocol, these new blocks would first need to be parsed into the format described in Sections~\ref{sec:systemdetails}. The new data would then be appended to the final rows of the appropriate weekly databases, with Address PIR DB data being subsequently sorted to maintain lexicographic ordering. 

After collecting a week's worth of new data in the weekly databases, these databases are subsequently emptied by having their data migrated to the monthly databases. Similarly, once a month's worth of new data is collected in the monthly databases, these databases are also emptied by having their data migrated to the all-time databases, where it permanently remains. This migration simply involves the new data being appended to the final rows of the appropriate monthly or all-time databases, with Address PIR DB data being subsequently sorted to maintain lexicographic ordering. In addition, whenever such updates occur, either when new data arrives in the weekly databases or when data is migrated, these changes are reflected in the manifest files of the corresponding databases.

This process ensures that the freshest data exists in the set of weekly databases, followed by the monthly and all-time databases. In addition, PIR servers do not need to perform synchronisation with each other with regards to the current state of the blockchain as that is implicitly handled by Bitcoin.

Our protocol can be further extended by having finer-grained temporal slices of
the Bitcoin blockchain, such as for the past day or hour. Having too many
temporal partitions, however, can expose the client to a privacy attack. A malicious PIR server can monitor a client's query and determine that the client is interested in hourly data, thus reducing the set of data in which that clients transaction of interest is included, when compared to the set of weekly data. This increases the likelihood of the malicious PIR server determining which addresses the client controls. 

\subsubsection{Integration with Bitcoin}
Our protocol as described in this paper can be easily integrated with Bitcoin and other cryptocurrencies with a similar SPV model. Our protocol acts as an alternative SPV implementation which is fully private, and as such would only require changes at the network protocol layer. At a minimum, new network message headers would need to be created which would allow clients to identify which full nodes support PIR SPV\@. Full nodes which do support PIR SPV would also need to account for extra storage to accommodate the additional databases and corresponding manifest files that are necessary to facilitate PIR based queries.
\end{document}